%% file: Hawk.tex
\documentclass[11pt]{article}
\usepackage{fullpage}

\usepackage{amsmath,amssymb}
\usepackage{hyperref}
\usepackage{graphicx}

\usepackage[utf8]{inputenc}
\DeclareUnicodeCharacter{00A0}{~}

\begin{document}

\input{defs}

\input{title}

\input{hawknum}

\section*{Acknowledgements}
I would like to thank R.~Casadio for helpful discussions and his encouragement to publish this trivial calculation. Moreover, I am grateful to A.~Alonso-Serrano for advising me of Ref.~\cite{Alonso-Serrano:2015bcr}.
This research is supported in part by the INFN, research initiative STEFI.

%\bibliographystyle{utphys}
%\bibliography{hawk}
\input{Hawk.bbl}
\end{document}

%% file: defs.tex
%  definitions

% i.e. and e.g.
\newcommand{\ie}{i.e.,\ }
\newcommand{\eg}{e.g.,\ }

% const
\newcommand{\const}{\operatorname{const.}} 

% sgn
%\newcommand{\sgn}{\operatorname{sgn}} 

% differentials (roman d)
\newcommand{\rmd}{\,\mathrm{d}}

% trace
\newcommand{\Tr}{\operatorname{tr}}

% Re and Im
\newcommand{\re}{\operatorname{Re}}
\newcommand{\im}{\operatorname{Im}}

% base of exponentials (roman e), with argument 
\newcommand{\e}[1]{\operatorname{e}^{#1}}

% life-time of black hole
\newcommand{\tBH}{t_{\text{ev}}}

%% file: title.tex
%\begin{titlepage}
\begin{center}

{\Large \textbf{Hawking radiation is corpuscular}}\\[1em]

\renewcommand{\thefootnote}{\fnsymbol{footnote}}
Wolfgang M{\"u}ck${}^{a,b}$\footnote[1]{wolfgang.mueck@na.infn.it}\\%
\renewcommand{\thefootnote}{\arabic{footnote}}
${}^a$\emph{Dipartimento di Fisica ``Ettore Pancini'', Universit\`a degli Studi di Napoli "Federico II"\\ Via Cintia, 80126 Napoli, Italy}\\
${}^b$\emph{Istituto Nazionale di Fisica Nucleare, Sezione di Napoli\\ Via Cintia, 80126 Napoli, Italy}\\[2em]

\abstract{The total number of Hawking quanta emitted during the evaporation of a Schwarzschild black hole is proportional to the square of the initial mass or, equivalently, to the Bekenstein entropy. This simple, but little appreciated, fact is interpreted in terms of the recent discovery of black hole soft hair.}

\end{center}
\vspace{2em}

%\paragraph{Keywords:} 
%\end{titlepage}

%\tableofcontents

%% file: hawknum.tex
\section{Estimate of the number of Hawking particles}
\label{sec1}

In 1974, Hawking discovered that black holes emit radiation with a thermal power spectrum \cite{Hawking:1974rv, Hawking:1974sw}. Soon after, Page gave an estimate of the life-time of a black hole from the knowledge of the emitted power \cite{Page:1976df}. In a simplified version of this calculation, one assumes that the black hole, at any value of the temperature, can be considered as an ideal black body emitting radiation in the form of massless bosons. As the black hole evaporates, its temperature is assumed to increase adiabatically as a function of the remaining mass. This simplification omits the grey-body factors \cite{Bekenstein:1977}, superradiance, and much of the backreaction of the radiation on the geometry. 

The purpose of this short note is to provide an estimate of the total numer of quanta emitted by the black hole. That number turns out to be proportional to the square of the black hole's initial mass in Planck units, which is reminiscent of the proposed graviton number in the black hole quantum $N$-portrait \cite{Dvali:2011aa, Dvali:2012en}. The simplicity of the calculation suggests that the result may be known, but I am not aware of it in the literature.\footnote{While this manuscript was under review, I learned from A.~Alonso-Serrano that the calculation was performed recently \cite{Alonso-Serrano:2015bcr}.}

The spectral luminosity density of an ideal black body is derived from Planck's formula \cite{Planck:1901} and reads\footnote{
Units are such that $\hbar=G=k_B=c=1$.}%
\begin{equation}
\label{lum.spec.density}
	\frac{dL}{d\omega} = \frac{gA}{8\pi^2}\frac{\omega^3}{\e{\omega/T}-1}~,
\end{equation}
where $A$ is the surface area, $T$ the temperature and $g$ the number of radiating degrees of freedom.\footnote{E.g., $g=1$ for a spinless boson, while $g=2$ for photons and gravitons.} 
The luminosity is obtained integrating \eqref{lum.spec.density} over the energy $\omega$,\footnote{For photons, this is the Stefan-Boltzmann law.}
\begin{equation}
\label{stefan.boltzmann}
	L = \frac{g \pi^2}{120} A T^4~.
\end{equation}

Instead of the luminosity, one may also consider the emission rate (number flux) of the emitted quanta. The number flux density is related to the luminosity density by
\begin{equation}
\label{num.flux.spec.density}
	\frac{d\Gamma}{d\omega} = \frac1{\omega}\frac{dL}{d\omega}= \frac{gA}{8\pi^2}\frac{\omega^2}{\e{\omega/T}-1}~.
\end{equation}
Integrating \eqref{num.flux.spec.density} over the energy, one finds the emission rate
\begin{equation}
\label{num.flux}
	\Gamma = \frac{g \zeta(3)}{4\pi^2} A T^3~,
\end{equation}
where $\zeta(x)$ denotes the Riemann zeta function.
%The mean energy of the quanta emitted at the temperature $T$ is\footnote{\eqref{mean.energy.T} resembles Wien's displacement law, which carries a different numerical factor.}
%
%\begin{equation}
%\label{mean.energy.T}
%	\bar{\varepsilon}(T) = \frac{L}{\Gamma} = \frac{\pi^4}{30\zeta(3)} T~.
%\end{equation}
%

Now consider a Schwarzschild black hole of mass $M$, with Hawking temperature and horizon area given by
\begin{equation}
\label{schwarzschild}
	T = \frac{\kappa}{2\pi} = \frac1{8\pi M}~,\qquad A= 4\pi r_s^2 = 16\pi M^2~,
\end{equation}
respectively. Because the emitted power corresponds to the mass loss rate, \eqref{stefan.boltzmann} and \eqref{schwarzschild} can be combined into
\begin{equation}
\label{dM.dt}
	\frac{dM}{dt} = - L = - \frac{g}{15\cdot 2^{11}\pi M^2}~,
\end{equation}
and one easily obtains the estimate for the evaporation time of the black hole,
\begin{equation}
\label{t.ev}
	\tBH = \frac{5\cdot 2^{11}\pi}{g} M_0^3~.
\end{equation}
where $M_0$ is the inital mass. 

The emission rate of Hawking quanta is found by substituting \eqref{schwarzschild} into \eqref{num.flux},
\begin{equation}
\label{BH.num.flux}
	\Gamma = \frac{g\zeta(3)}{128\pi^4} \frac1{M}~.
\end{equation}
Since Page's paper, the emission rate has appeared repeatedly in the literature, see, \eg the recent papers \cite{Visser:2014ypa, Gray:2015pma} and references therein. It is very surprising that the total number of Hawking quanta emitted during the evaporation appears to have never been considered. Combining \eqref{BH.num.flux} with \eqref{dM.dt}, one finds
\begin{equation}
\label{dN.dM}
	\frac{dN}{dM} = \frac{\Gamma}{\frac{dM}{dt}} = -\frac{240\zeta(3)}{\pi^3} M~.
\end{equation}
A simple integration yields the total number of Hawking quanta
\begin{equation}
\label{N.BH}
	N = \frac{120\zeta(3)}{\pi^3} M_0^2~,
\end{equation}
which is equal, up to a numerical factor, to the initial Bekenstein entropy
\begin{equation}
\label{S.BH}
	N = \frac{30\zeta(3)}{\pi^4} S~,
\end{equation}
independently of the number of radiating degrees of freedom. 

The semi-classical analysis remains valid as long as $M\gg 1$. The quantum regime may be modelled by an ultraviolet cut-off on the energy and mass integrals and gives rise to $1/N$ corrections. Hence, for a macroscopic black hole, the final stage of the evaporation, which happens in a quantum gravity regime, appears to be irrelevant in terms of the emitted Hawking quanta and of the information they carry.
The inclusion of greybody factors will likely affect the numerical prefactor in \eqref{N.BH} and \eqref{S.BH}.

\section{Conclusions}
\label{sec2}

The above results appear to be relevant for recent developments on the physics of black holes and, in particular, for the information paradox. First, \eqref{N.BH} is recognized as the proposed number of gravitons in the black hole quantum $N$-portrait \cite{Dvali:2011aa, Dvali:2012en}. In this proposal, a black hole is pictured as a Bose-Einstein condensate of a very large number, $N\sim M^2$, of gravitons at the verge of a quantum phase transition. Hawking radiation is the result of the depletion of the condensate caused by 2-body interactions and contains non-thermal features of order $1/N$, which resolve the information paradox \cite{Dvali:2012rt, Dvali:2015aja}. The results \eqref{N.BH} and \eqref{S.BH} provide direct evidence in support of the quantum $N$-portrait, in the sense that a semi-classical black hole may be considered as a bound state of the $N$ Hawking particles it dissolves into. Interpreting each emitted Hawking particle as an information-carrying unit, \eqref{S.BH} is indeed the expected relation between the entropy and the particle number. This indicates that it is more appropriate to give Hawking radiation, which is sparse in the semi-classical regime \cite{Gray:2015pma}, a corpuscular interpretation instead of an undulatory one.

Secondly, the results should be interpreted in the light of recent developments on the infrared structure of quantum gravity, which revived  results of Bondi, van der Burg, Metzner and Sachs (BMS) \cite{Bondi:1962px, Sachs:1962wk} from the early 1960s. Soft symmetries, or better, the \emph{supertranslations} of the extended BMS symmetry group \cite{Barnich:2009se, Barnich:2011mi}, give rise to conserved charges \cite{Strominger:2013jfa, He:2014laa, Strominger:2014pwa, Kapec:2014opa, Flanagan:2015pxa} of the gravitational $S$-matrix. These soft supertranslation charges constitute what has been known as gravitational memory. Extending these results to the asymptotic symmetries of 
the black hole horizon \cite{Donnay:2015abr, Dvali:2015rea, Hotta:2016qtv}, it has been discovered that black holes carry `soft hair' \cite{Hawking:2016msc}, which retains the information about the state before the black hole formation and imprints that information, as the black hole evaporates, on the outgoing Hawking radiation. Classically, the information storage capacity of the horizon is infinite, but it is physically impossible to excite soft quanta that are smaller than the Planck area on the horizon, giving rise to an effective pixelization in agreement with the Bekenstein entropy-area law. \eqref{N.BH} and \eqref{S.BH} provide evidence for this in the evaporation process. The information about the final state is measurable by the gravitational memory effect, \ie the soft supertranslation charges at future null infinity that are carried by the outgoing Hawking quanta. This points again at the importance of the corpuscular point of view of Hawking radiation. The effective, semi-classical, gravitational memory capacity used by the Hawking quanta is of order $N\sim M_0^2$. Hence, it appears that unitarity is preserved, after all, when a gravitational system of total mass $M_0$ collapses into a black hole and evaporates subsequently.

%% file: Hawk.bbl
\providecommand{\href}[2]{#2}\begingroup\raggedright\endgroup